\documentclass[12pt]{article}
\usepackage{amsmath,epsfig,amssymb,eufrak}
\usepackage{amscd,latexsym}
\usepackage{color}

\date{}

\def\lsi{\raise0.3ex\hbox{$<$\kern-0.75em\raise-1.1ex\hbox{$\sim$}}}
\def\gsi{\raise0.3ex\hbox{$>$\kern-0.75em\raise-1.1ex\hbox{$\sim$}}}

\newcommand{\beq}{\begin{equation}}
\newcommand{\eeq}{\end{equation}}
\newcommand{\beqn}{\begin{eqnarray}}
\newcommand{\eeqn}{\end{eqnarray}}

\newcommand{\bchi}{\bar\chi}

\input macros.sty      
\input eqalign.sty     

\begin{document}

\begin{titlepage}

\title{
  {\vspace{-0cm} \normalsize
  \hfill \parbox{40mm}{DESY/05-206\\
                       October 2005}}\\[10mm]
Two--dimensional lattice Gross--Neveu model\\
with Wilson twisted mass fermions}
\author{Kei-ichi Nagai and Karl Jansen          
\\
\\
{\small   John von Neumann-Institut f\"ur Computing NIC,} \\
{\small         Platanenallee 6, D-15738 Zeuthen, Germany} \\ \ \\
}

\maketitle

\begin{abstract}
We study the two-dimensional lattice Gross--Neveu model
with Wilson twisted mass fermions 
in order to explore the phase structure in this setup. 
In particular, we investigate the behaviour of the phase transitions
found earlier with standard Wilson fermions as a function of the 
twisted mass parameter $\mu$. We find that qualitatively the dependence 
of the phase transitions on $\mu$ is very similar to the 
case of lattice QCD.
  \vspace{0.75cm}
\noindent
\end{abstract}

\end{titlepage}

\section{Introduction}
\label{sec:intro}

Recent results from simulations of lattice QCD with Wilson twisted 
mass fermions \cite{Frezzotti:2000nk,Frezzotti:2003ni} 
revealed an 
intriguing phase structure 
with an Aoki phase at strong coupling \cite{Ilgenfritz:2003gw,Sternbeck:2003gy} 
and pronounced first order phase transitions at intermediate 
values of the coupling which 
weakens when the continuum limit is approached 
\cite{Farchioni:2004us,Farchioni:2004ma,Farchioni:2004fs,Farchioni:2005tu}.  
These findings are supported by calculations in chiral perturbation 
theory for standard Wilson fermions \cite{Sharpe:1998xm} which were 
later extended to the case of Wilson twisted mass fermions
\cite{Munster:2004am,Scorzato:2004da,Sharpe:2004ny,Sharpe:2004ps,Aoki:2004ta,Sharpe:2005rq}.
See refs.~\cite{Frezzotti:2004pc,Shindler} for reviews.
The appearance of the first order phase transition turns out to be a 
generic phenomenon of Wilson fermions and takes place independent of 
whether the twisted 
mass parameter is switched on or not. 

Of course, one may argue that on a finite lattice no phase transition 
can occur and that the metastabilities observed in 
refs.~\cite{Farchioni:2004us,Farchioni:2004ma,Farchioni:2004fs,Farchioni:2005tu} 
are purely an artefact of the algorithm employed. 
From the analytical side, 
chiral perturbation theory ($\chi$PT) is a powerful tool
to determine the phase structure. 
However, the assumption of $\chi$PT is that $a \Lambda \ll 1$,
where $a$ is the lattice spacing and $\Lambda$ is the scale parameter.
Moreover, the analysis by $\chi$PT is done up to $O(a^2)$ only.
Thus, it is unclear whether at large values of $a$ and by inclusion 
of higher order terms the picture can change substantially.

Motivated by these considerations and 
in order to have another angle on the problem, we decided
to study the phase structure of the Gross-Neveu model \cite{Gross:1974jv} 
on the lattice including a twisted mass term, 
taking this 
as a toy model of twisted mass lattice QCD.
In the large-$N$ limit, where $N$ is the number of flavours,
the Gross-Neveu (GN) model is asymptotically free and has bound states 
\cite{Gross:1974jv}.  
The GN-model is one-loop exact in the large-$N$ limit
and hence, in this limit, higher order contributions of the lattice spacing 
are taken into account.
In this sense, the large-$N$ investigation of the 
GN-model goes beyond $\chi$PT and can provide therefore a complementary 
picture of the phase diagram, although one should always keep in mind
that we are considering a two-dimensional model only.

The GN-model has been studied 
for pure Wilson fermions 
in the large-$N$ limit 
in refs.~\cite{Izubuchi:1998hy,Aoki:1985jj}.
In ref.~\cite{Aoki:1983qi}, the phase structure of the lattice 
GN-model with Wilson fermions  
is first predicted and a 
so-called {\it Parity-Flavour broken phase} (or {\it Aoki phase}) was 
found. A refined analysis \cite{Izubuchi:1998hy} showed that even the 
zero temperature phase diagram is rather complex and reveals 
besides the second order phase transition to the Aoki phase 
strong first order phase transitions. 
In this work we want to extend the investigations of 
refs.~\cite{Aoki:1983qi,Aoki:1985jj,Izubuchi:1998hy} to the case when 
a twisted mass parameter $\mu$ is switched on. In particular,
we want to address the question
of the $\mu$-dependence
of the first order phase transition.
We hope that our findings can give some, at least qualitative insight also 
into the generic phase diagram of lattice QCD with Wilson fermions.

\section{Lattice Gross--Neveu model with twisted mass fermion}
\label{sec:lat}


A formulation of the two-dimensional Gross--Neveu model on the lattice 
with Wilson fermions
is given in refs.~\cite{Aoki:1983qi,Aoki:1985jj,Izubuchi:1998hy}.
Our study here follows  
refs.~\cite{Aoki:1985jj,Izubuchi:1998hy} where the 
the lattice couplings of the model are treated separately.
We choose the so-called  twisted basis
\cite{Frezzotti:2003ni} for our investigation in which the Lagrangian 
of the model reads
\beq
{\cal L}_{GN}=\bchi \left[\gamma \cdot \nabla^{sym} -
r \nabla_\mu \nabla_\mu^\dagger + M_{cr}(r) 
+ m_q e^{i \gamma_5 \tau_3 \omega} \right] \chi
-\frac{g_\sigma^2}{2 N}  (\bchi \chi)^2 
-\frac{g_\pi^2}{2 N}  (\bchi i \gamma_5 \tau_3\chi)^2 ,
\label{eq:lat-gn-chi}
\eeq
where $(\bchi,\chi)$ denote the fermion fields in the ``twisted basis''
or $\chi$-basis,
and $N$ is the number of the ``flavours''. 
The parameters of the model are the critical fermion mass 
$M_{cr}$, the off-set fermion mass $m_q$, the twist angle 
$\omega$ and the couplings $g_\sigma$ and $g_\pi$.
In this version of twisted mass fermions, the Wilson term is not twisted.
An important remark is that in general the coupling constant 
$g_\sigma \ne g_\pi$ 
on the lattice\cite{Aoki:1985jj,Izubuchi:1998hy}.

In momentum space, the Lagrangian in the $\chi$-basis is given as
\beq
{\cal L}_{GN}=\bchi \left[K(p) + W(p) + m_0 
+ i \gamma_5 \tau_3 \mu  \right] \chi
-\frac{g_\sigma^2}{2 N}  (\bchi \chi)^2 
-\frac{g_\pi^2}{2 N}  (\bchi i \gamma_5 \tau_3  \chi)^2 ,
\label{eq:gn-chi-mom}
\eeq
where $K(p)=i \sum_\nu \gamma_\nu \sin (p_\nu a)$,
$W(p)=r \sum_\nu (1-\cos(p_\nu a))$,
$m_0 = M_{cr} + m_q \cos \omega$,
$\mu = m_q \sin \omega$
and
$r$ the Wilson parameter.
The situation of ``full twist'' is realized when $\omega=\pm \pi/2$ 
or, equivalently, $m_0=M_{cr}$. 

In the  
large-$N$ limit we obtain the Lagrangian
\be
{\cal L}_{GN}= 
\bchi \left[K(p) + W(p) +
\sigma_L
+ i \gamma_5 \tau_3 \pi_L  \right] \chi 
+\frac{N}{2 g_\sigma^2}  (\sigma_L- m_0)^2
+\frac{N}{2 g_\pi^2}  (\pi_L - \mu)^2 ,
\label{eq:gn-chi-1}
\eeq
where, by the equations of motion, the bosonic fields are defined as
\beq
\sigma_L = - \frac{g_\sigma^2}{N} ( \bchi \chi) + m_0 \quad, \quad 
\pi_L =-\frac{g_\pi^2}{N} (\bchi i \gamma_5 \tau_3  \chi )+ \mu \quad.
\eeq

To solve the model, we integrate out the fermion fields, treating 
the $\pi$- and $\sigma$-fields as constant since 
the quantum fluctuations are suppressed in the large-$N$ limit.
We then obtain the effective potential
$V_{eff}$,
\beq
Z=\int [d \chi][d \bchi] \exp \{- \int d^2 x {\cal L}_{GN}\} 
\equiv \exp \{ -N \Omega V_{eff} \} \, ,
\eeq
where $\Omega$ is the space-time volume.

\noindent $V_{eff}$ in large-$N$ limit is given by       
(setting here and in the following $r=1$),
\beqn
\hspace{-1cm}
&&V_{eff}=\frac{1}{2 g_\sigma^2}(\sigma_L - m_0)^2
+\frac{1}{2 g_\pi^2}(\pi_L - \mu)^2 
\label{eq:eff-pot-chi} \\
\hspace{-1cm}
&& - \int_p \ln [ \sum_\nu \sin^2 (p_\nu a) 
+\{\sum_\nu (1 -\cos (p_\nu a)) \}^2 \nonumber 
 +(\sigma_L^2+\pi_L^2) 
+ 2 \sigma_L
 \sum_\nu (1 -\cos (p_\nu a))] ,
\nonumber 
\eeqn
where $\int_p=\int_{-\pi}^\pi \frac{d^2(p a)}{(2 \pi)^2}$.
Note that eq.~(\ref{eq:eff-pot-chi}) has the same form
as the effective potential  
in the Wilson fermion formulation of 
the GN-model on the 
lattice \cite{Aoki:1985jj,Izubuchi:1998hy}.
The difference shows up only in the definition of the 
field $\pi_L$, where there is an explicit appearance 
of the twisted mass parameter $\mu$.  

\subsection{Phase structure and Pion mass in twisted basis}
\label{sec:phase}

At $\mu=0$, i.e. for pure Wilson fermions, the phase structure 
of the Gross-Neveu model is 
already quite complicated showing
first order and 
second order phase transitions. 
It is unclear, whether the presence of the twisted mass term 
can change the phase structure.
In order to explore the phase structure of the lattice GN-model, 
it is sufficient to solve 
the saddle point equations as they can be derived
from the effective potential $V_{eff}$.

In the following we will
concentrate on the twisted basis,
because then the effective potential 
has the same form as the 
one for
ordinary Wilson fermion.
The effect of the twisted mass parameter is, to 1-loop order, 
only included 
via the field $\pi_L$ by a shift. 

The saddle point equations are 
\beqn
&&\frac{\partial V_{eff}}{\partial \sigma_L}
= \frac{\sigma_L - m_0}{g_\sigma^2} - \sigma_L F(\sigma_L,\pi_L)
 - G(\sigma_L,\pi_L) = 0, 
\label{eq:saddle-s} \\
&&\frac{\partial V_{eff}}{\partial \pi_L }
= \frac{\pi_L - \mu}{g_\pi^2} - \pi_L F(\sigma_L,\pi_L)=0,
\label{eq:saddle-p} 
\eeqn
where
\beqn
F(\sigma_L,\pi_L) &=& \int_p \frac{2}{A(\sigma_L,\pi_L)} , \quad
G(\sigma_L,\pi_L) = \int_p \frac{2 \sum_\nu (1-\cos (p_\nu a))}{A(\sigma_L,\pi_L)} , \nonumber \\
 A(\sigma_L,\pi_L) &=& (\sigma_L^2+\pi_L^2) 
+ \{\sum_\nu (1-\cos (p_\nu a)) \}^2 + \sum_\nu \sin^2 (p_\nu a) \nonumber \\
& & + 2 \sigma_L \sum_\nu (1-\cos (p_\nu a)) \, . \nonumber
\label{eq:func}
\eeqn

In the saddle point equations 
the 1-loop tadpole contributions are canceled 
by the potential terms of $\sigma_L$ and $\pi_L$.
The solution of the saddle point equations
is non-trivial and will differ from the one obtained with 
pure Wilson fermions opening thus the possibility of new
phenomena.

The neutral pion mass and the sigma particle mass are computed 
from the second derivative of the effective potential at the 
solution of the saddle point equations,
\beqn
\hspace{-1cm}
&&M_\pi^2 = \frac{1}{2} \frac{\partial^2 V_{eff}}{\partial \pi_L^2}
= \frac{1}{2} \left[ \frac{1}{g_\pi^2} - F(\sigma_L,\pi_L)
- \pi_L \frac{\partial}{\partial \pi_L} F(\sigma_L,\pi_L)  \right] \, ,
\label{eq:pionmass1}  \\
\hspace{-1cm}
&&M_\sigma^2 = \frac{1}{2} \frac{\partial^2 V_{eff}}{\partial \sigma_L^2}
= \frac{1}{2} \left[ \frac{1}{g_\sigma^2} - F(\sigma_L,\pi_L)
- \sigma_L \frac{\partial}{\partial \sigma_L} F(\sigma_L,\pi_L)
- \frac{\partial}{\partial \sigma_L} G(\sigma_L,\pi_L) \right] .
\label{eq:sigmamass1} 
\eeqn

Our strategy is the following. We first fix the values 
the couplings $g_\sigma^2$ and $g_\pi^2$ such  
that the perturbative tuning 
relation between 
$g_\pi^2$ 
and  $g_\sigma^2$ as computed for the ordinary Wilson fermion formulation 
\cite{Izubuchi:1998hy} holds, 
\beq
\frac{1}{g_\pi^2} = \frac{1}{g_\sigma^2} + 4 C_2 + O(a) \, , 
\quad \quad C_2=\frac{2 \sqrt{3}}{27}+\frac{1}{12 \pi} \, .
\label{tuning}
\eeq
In the following table,
the (perturbative) tuning values for various $g_\sigma^2$ are given.

\begin{center}
\begin{tabular}[t]{|c|c|c|c|c|}
\hline
$g_\sigma^2$ & 0.1   & 0.5   & 1.0  & 2.0 \\
\hline
$g_\pi^2$    & 0.094 & 0.381 & 0.617& 0.893 \\
\hline 
\end{tabular}
\end{center}

\vspace{0.5cm}

We then also fix the value of 
the twisted mass $\mu$ and 
determine 
the values of $\sigma_L$ and $\pi_L$  
from the solutions of the saddle point equations 
eqs.~(\ref{eq:saddle-s}, \ref{eq:saddle-p}) as a function of $m_0$.
In this way, we can study whether $\sigma_L$ or $\pi_L$ undergo    
a phase transition. Once we have the value for $\sigma_L$ and $\pi_L$ 
we can insert them into the gap equations, 
eqs.~(\ref{eq:pionmass1}, \ref{eq:sigmamass1}) to compute also the 
corresponding masses.

Let us remark that
in principle there is also an additional        
tuning relation between the coupling 
$g_\sigma^2$ and the fermion mass $m_0$. To explore the phase structure, 
we will, however, vary $m_0$ and hence will not make use of 
this tuning relation.
We will also restrict ourselves to values of $g_\sigma^2 \le 2$, since for
larger values of the coupling the tuning relation of 
eq.~(\ref{tuning}) will not hold anymore.

\section{Results}
\label{sec:results}

In this section we will give our results obtained 
from the solution of the saddle point 
equations. We used a lattice of 
size $128^2$ and solved the equations numerically. We always checked that
we indeed found the correct solutions by inspecting the extrema of 
the effective 
potential itself. 

\begin{figure}[h!tb] 
\centering
\vspace*{-0.8cm}
\centerline{
\resizebox{9.0cm}{!}{\rotatebox{-90}{\includegraphics{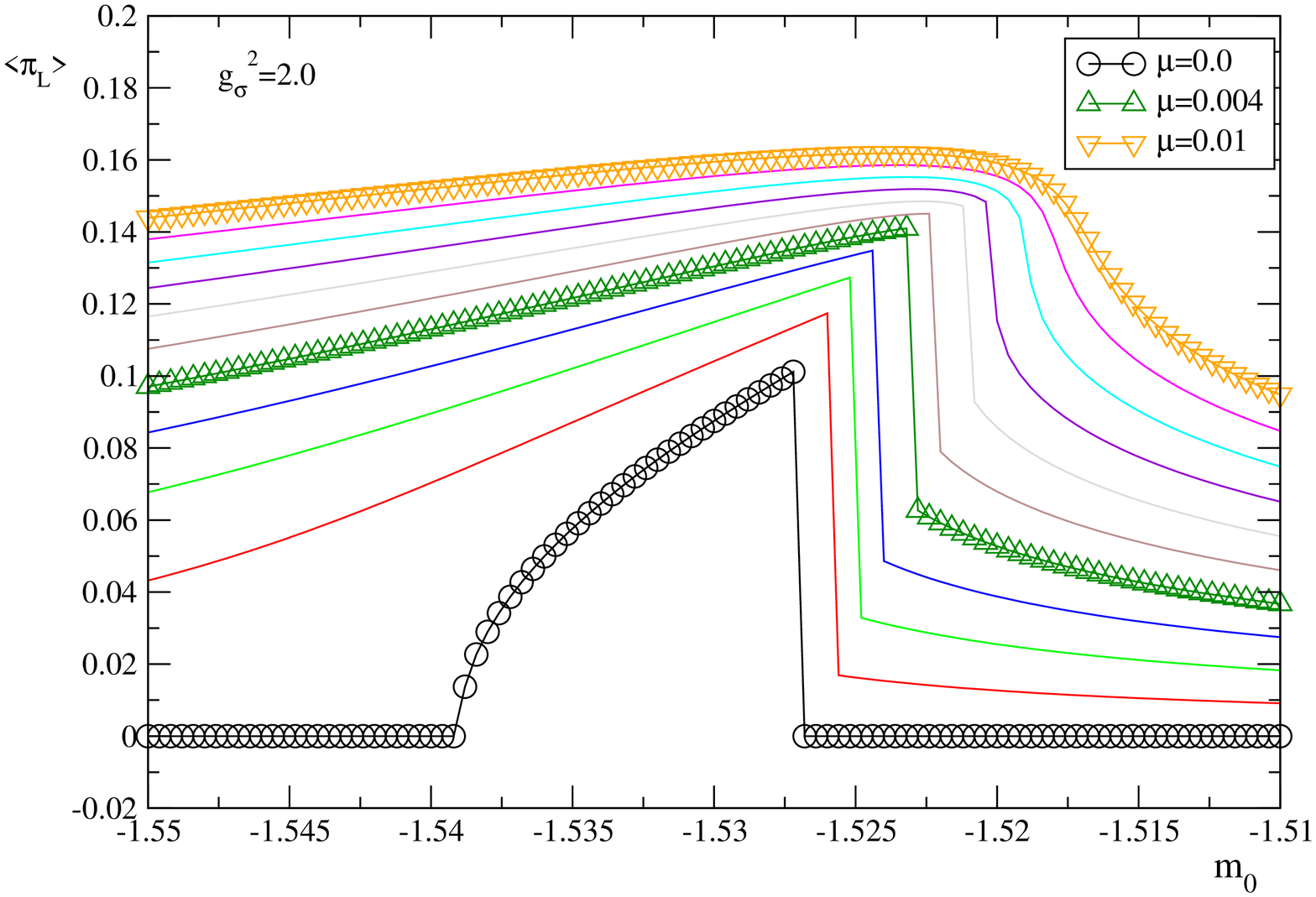}}}
}
\vspace*{-0.6cm}
\centerline{
\resizebox{9.0cm}{!}{\rotatebox{-90}{\includegraphics{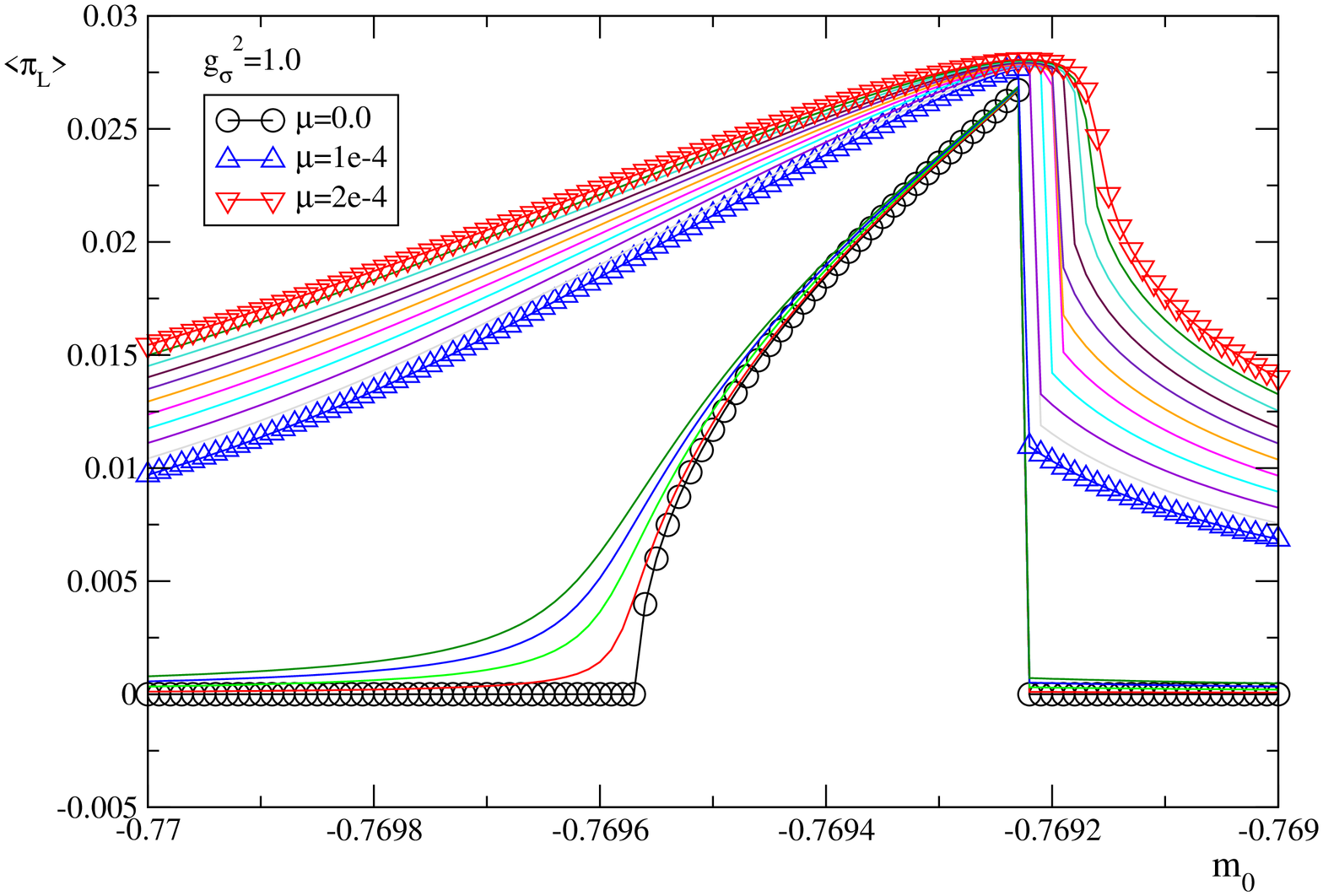}}}
}
\vspace*{-0.6cm}
\centerline{
\resizebox{9.0cm}{!}{\rotatebox{-90}{\includegraphics{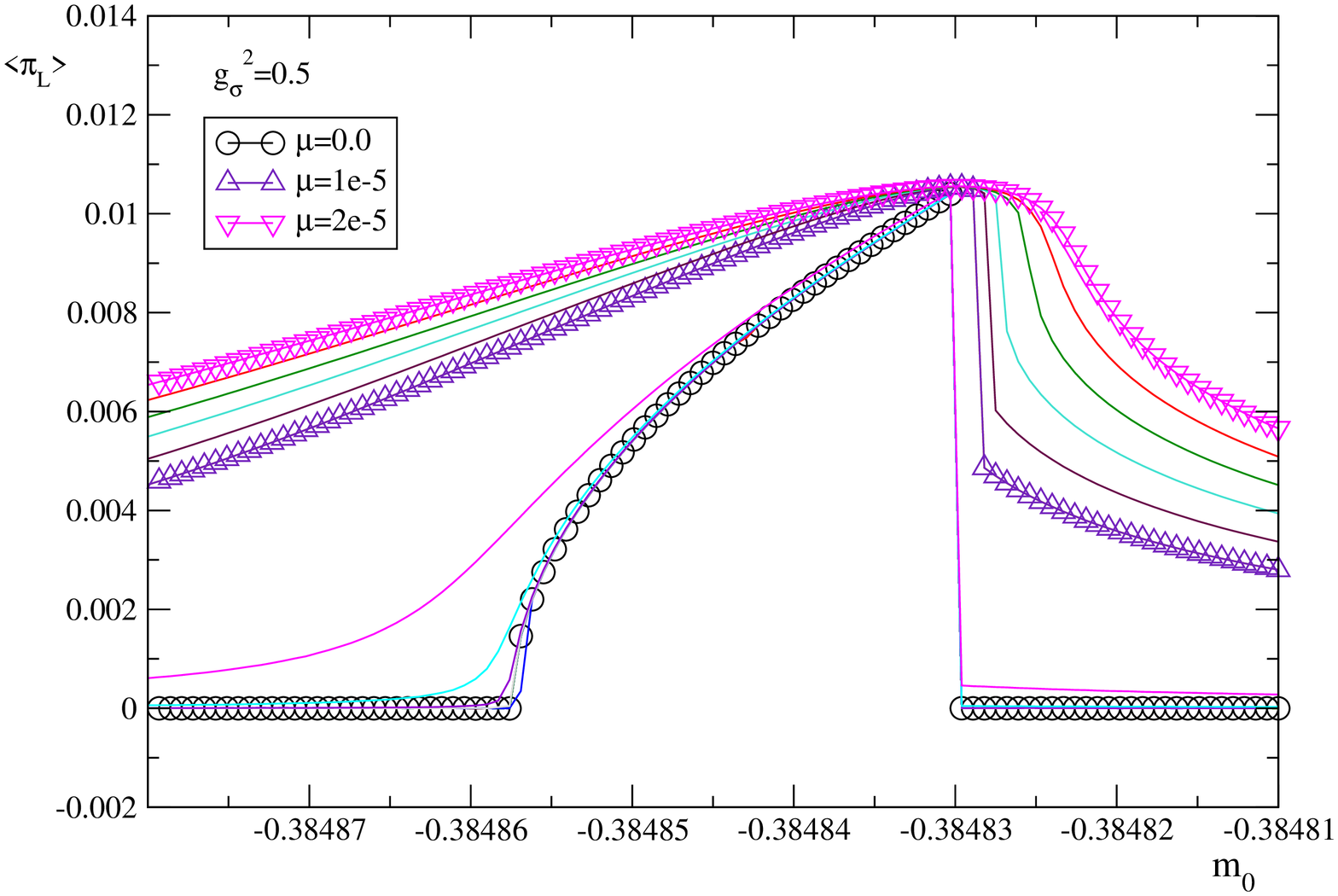}}}
}
\caption{$\langle \pi_L \rangle$ as a function of 
the fermion mass $m_0$. The coupling $g_\pi$ is the perturbatively 
tuned one obtained from eq.~(\ref{tuning}) and 
assumes values of $g_\pi^2= 0.893$, $g_\pi^2= 0.617$ and $g_\pi^2= 0.381$ 
from top to bottom. 
We represent some values of the twisted mass parameter $\mu$ by open 
symbols. Other, intermediate values of $\mu$ are represented
by solid lines in the graph.
}
\label{fig:exppi}
\end{figure}

In fig.~\ref{fig:exppi} we show the expectation value 
$\langle\pi_L\rangle$ of the $\pi$-field as a function of the mass parameter 
$m_0$ for various values of the twisted mass parameter $\mu$. 
Let us summarize the results of our large-$N$ computations.
At $\mu=0$ and
increasing the values of $m_0$, we find 
in accordance with ref.~\cite{Izubuchi:1998hy} first   
and second order phase transitions, where the 
$\pi$-field assumes a non-zero value. For larger values of 
$m_0$ there is a first order 
transition where $\langle\pi_L\rangle$ jumps back to a zero value. 
This structure changes in two aspects when the twisted mass parameter 
$\mu$ is switched on. First, the second order phase transition vanishes 
and $\langle \pi_L \rangle$ shows an analytical behaviour. 
Second, for some range 
of $\mu < \mu_\mathrm{crit}$ the first order phase transition remains. 
However, it becomes weaker when $\mu$ is increased and turns into a second
order phase transition when a critical value $\mu_\mathrm{crit}$ is reached. 
For $\mu > \mu_\mathrm{crit}$ the phase transition disappears altogether
and turns into a completely smooth and analytical behaviour. 

When the coupling $g_\sigma^2$ is made smaller, with the associated
tuning of $g_\pi^2$, the region in $m_0$ where the 
phase transitions occur shrinks and also the jumps at the first order 
phase transitions become smaller.  
Thus the qualitative picture does not change when we approach the 
continuum limit. 
Note that also the value of $\mu_\mathrm{crit}$ gets smaller with 
decreasing coupling $g_\sigma^2$.

The fate of the first order phase transition 
is very much reminiscent of the situation of four-dimensional lattice QCD.
Also there it has been found from numerical simulations 
\cite{Farchioni:2004fs,Farchioni:2004ma,Farchioni:2004us}
and from calculations in chiral perturbation theory
\cite{Sharpe:1998xm,Sharpe:2004ny,Munster:2004am,Scorzato:2004da,Sharpe:2005rq}
that the first order phase transition terminates at a critical 
values of $\mu$. 
Note that the values of $m_0$ where the first order phase transition takes 
place moves to larger values when $\mu$ is increased. 
A similar phenomenon can be expected also in lattice QCD.
What is very interesting from our analysis in the GN-model 
--and apparently different from current findings
of lattice QCD and $O(a^2)$-$\chi$PT-- is that 
this first order phase transition is accompanied by a second 
order phase transition at $\mu=0$ or its remnant at $\mu>0$. 

\begin{figure}[h!tb] 
\centering
\centerline{
\resizebox{9.5cm}{!}{\rotatebox{-90}{\includegraphics{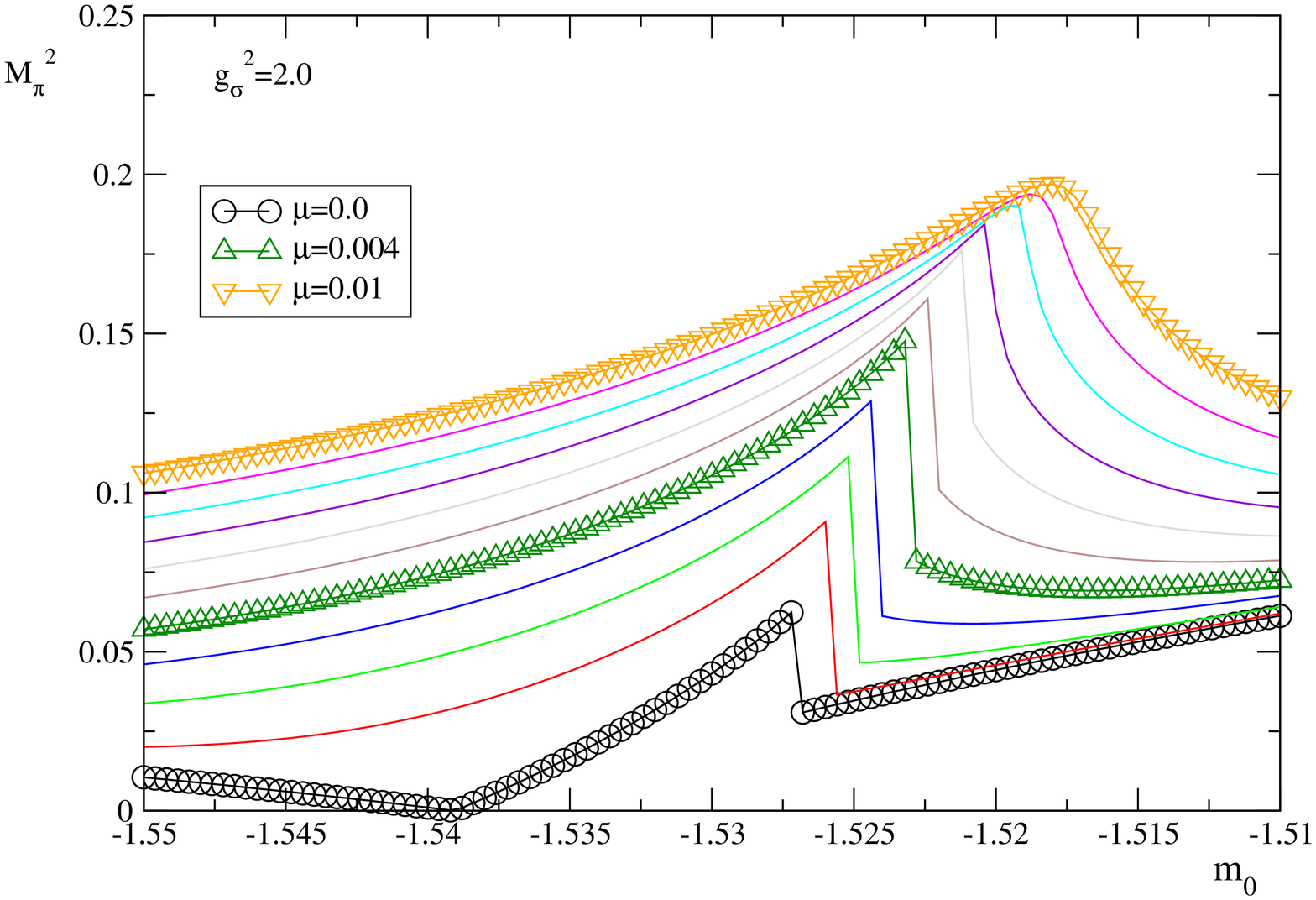}}}
}
\centerline{
\resizebox{9.5cm}{!}{\rotatebox{-90}{\includegraphics{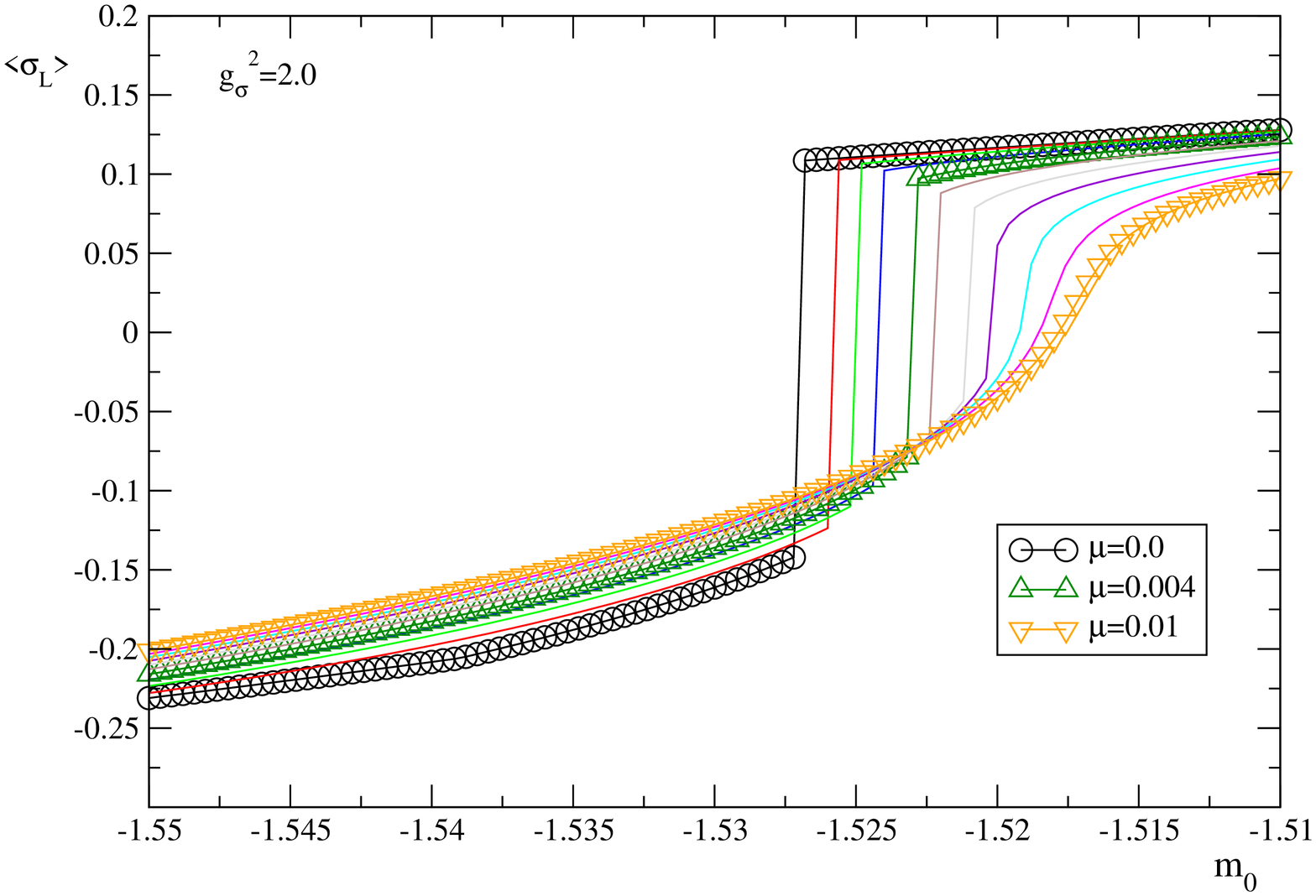}}}
}
\vspace*{-0.5cm}
\centerline{
\resizebox{9.5cm}{!}{\rotatebox{-90}{\includegraphics{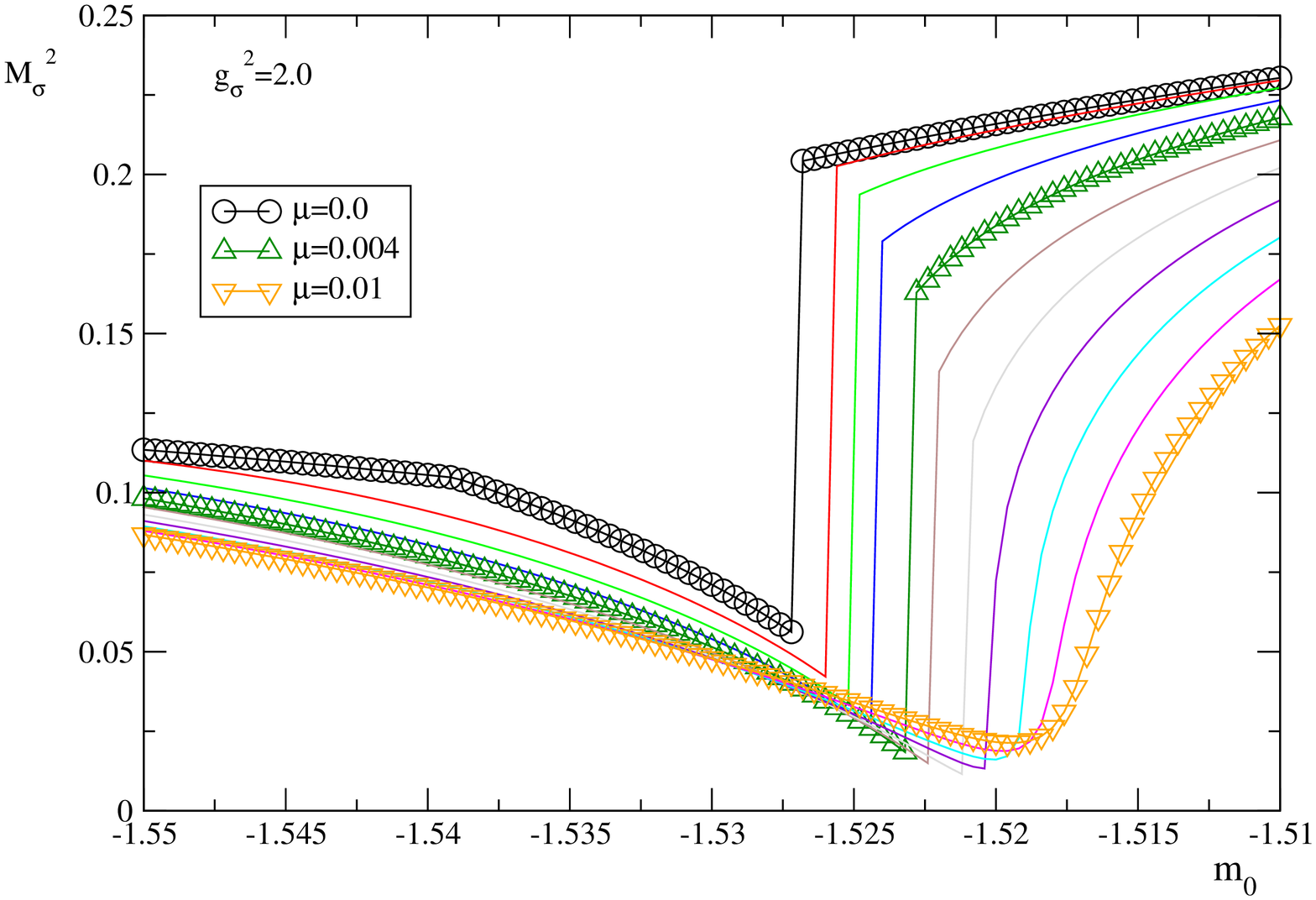}}}
}
\caption{
We show (from top to bottom) 
the pion mass, the $\sigma$-field expectation value and 
the $\sigma$-mass as a function 
of $m_0$ at $g_\sigma^2=2$, $g_\pi^2=0.893$. 
}
\label{fig:masspi}
\end{figure}

In fig.~\ref{fig:masspi} we show at the example of the coupling values
$g_\sigma^2=2$, $g_\pi^2=0.893$ the behaviour of the pion mass (top), the 
$\sigma$-field expectation value (middle) and the $\sigma$-mass (bottom). 
At $\mu=0$ and 
increasing $m_0$, the pion mass first goes to zero 
and vanishes at the second order phase transition. 
It then assumes non-zero values again until it jumps when the first order 
phase transition is hit. Switching on $\mu$, the second order phase transition 
vanishes and the behaviour of the pion mass
becomes continuous. The first order phase transition remains, however, 
until $\mu=\mu_\mathrm{crit}$. For larger values of $\mu$ 
also the first order phase transition vanishes and the pion mass shows
an overall analytical behaviour. 

The $\sigma$-field expectation value and mass do not show a 
large sensitivity on the second order phase transition. However, this 
field feels the first order phase transition strongly and shows jumps 
in the field expectation value as well as in the mass. The picture shown 
here for the couplings $g_\sigma^2=2$, $g_\pi^2=0.893$ stays qualitatively 
unchanged when the coupling $g_\sigma^2$ is reduced with the associated
tuning of $g_\pi^2$. 

\begin{figure}[h!tb]
\centering
\centerline{
\resizebox{9.5cm}{!}{\rotatebox{-90}{\includegraphics{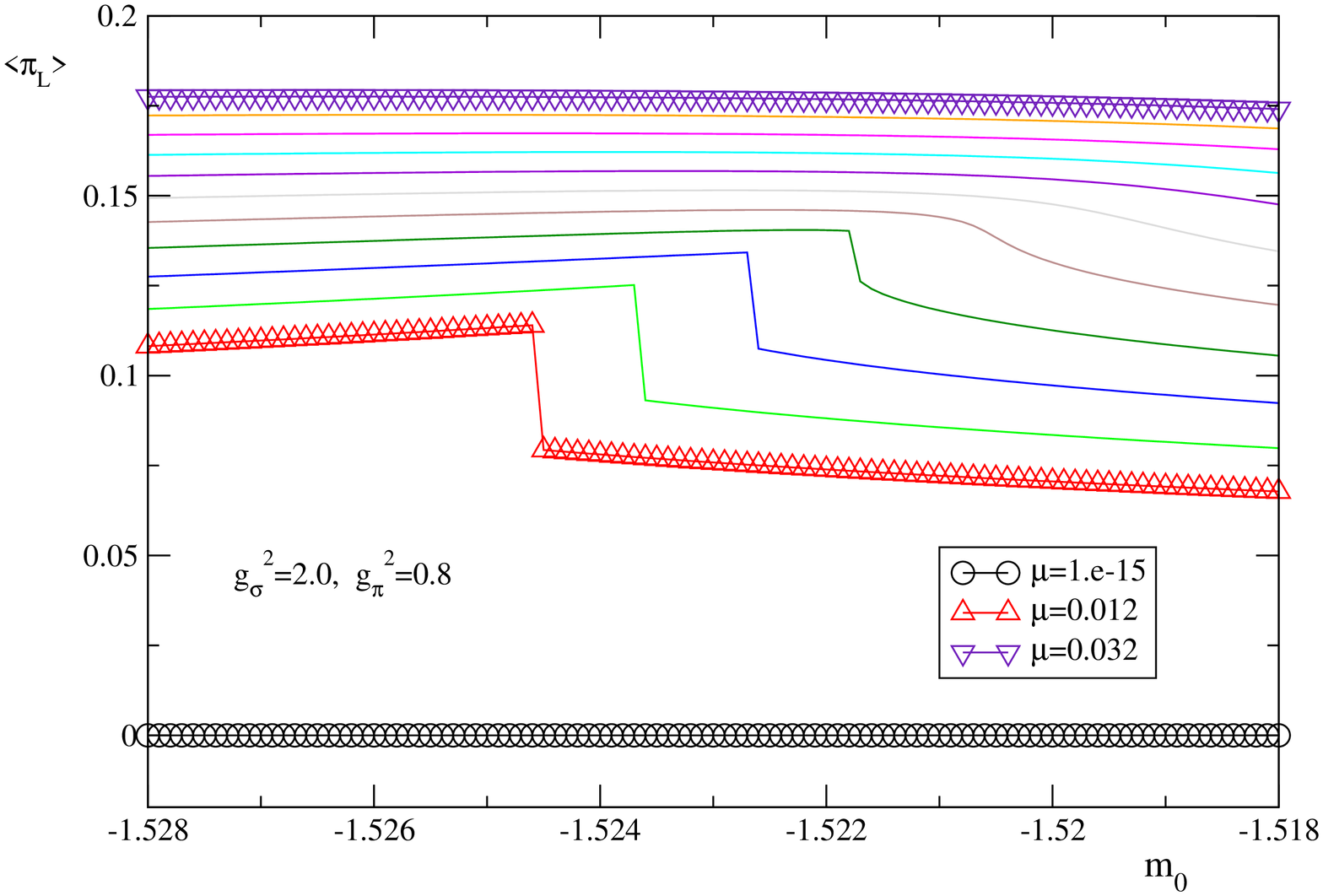}}}
}
\vspace*{-0.5cm}
\centerline{
\resizebox{9.5cm}{!}{\rotatebox{-90}{\includegraphics{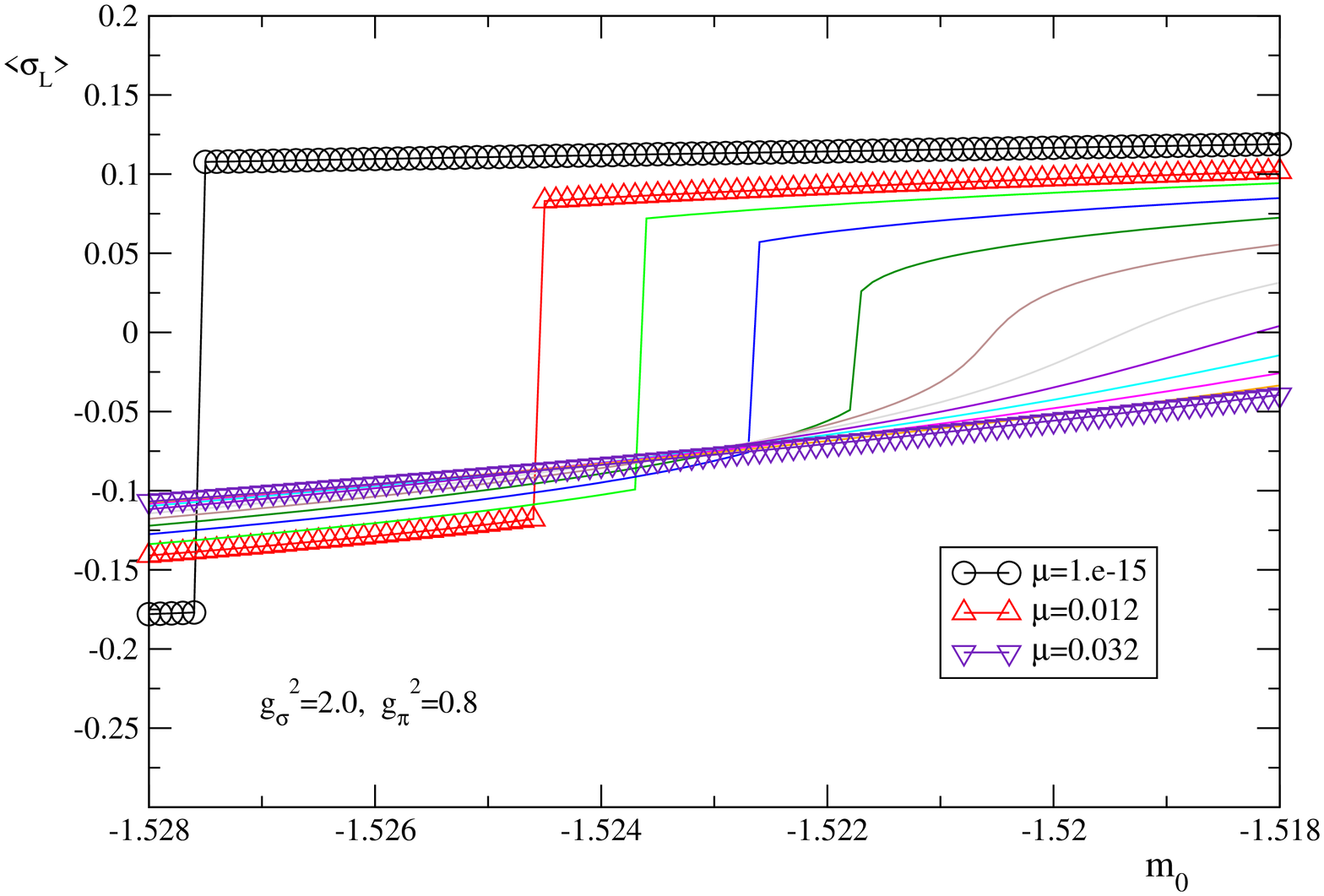}}}
}
\caption{
The $\pi$- (top) and $\sigma$-field (bottom) expectation values 
as a function of $m_0$ at the so-called $\sigma$-phase transition
\cite{Izubuchi:1998hy}.
}
\label{fig:sigmapt}
\end{figure}

Finally and for completeness, we show in fig.~\ref{fig:sigmapt} the fate 
of the so-called $\sigma$-phase transition \cite{Izubuchi:1998hy}
for a value of the coupling $g_\sigma^2=2$.
Note that the value of the coupling is now $g_\pi^2=0.8$ and does 
not correspond to the ones of the perturbative tuning.
This phase transition is characterized by the fact that at $\mu=0$ the 
$\sigma$-field expectation value $\langle \sigma_L \rangle $ shows a discontinuous jump, 
whereas the 
$\pi$-field expectation value stays essentially zero. 
This phase transition 
survives when $\mu$ is switched on and  
can clearly be seen in the jump of $\langle \sigma_L \rangle$. 
It is interesting to observe that the point in $m_0$ where this jumps
happens is moved to larger values of $m_0$ when $\mu$ is increased. 
However, also for this case the phase transition vanishes when 
$\mu$ is increased above a certain critical value.  
Furthermore, the $\pi$-field shows now at $\mu>0$ also a jump and 
exhibits the first order phase transition.

\section{Summary}
\label{sec:summary}

Following the earlier work of 
refs.~\cite{Aoki:1983qi,Aoki:1985jj,Izubuchi:1998hy}  
we studied in this letter, how the presence of the twisted mass 
term influences the known phase diagram of the 
two-dimensional  Gross--Neveu model which we consider as a 
toy model of lattice QCD. 
We found that the coexistence of the Aoki phase and the 
first order phase transition as it is obtained 
at vanishing twisted mass parameter $\mu$ \cite{Izubuchi:1998hy} 
is {\em not} kept 
at $\mu>0$. 
The first change is that at $\mu>0$ the second order phase transition 
turns into a smooth and analytical behaviour, although e.g. the 
$\pi$-field expectation $\langle \pi_L \rangle$ still shows a rapid change. 
The second change is that the gap in $\langle \pi_L \rangle$ of 
the first order phase transition shrinks with increasing values of $\mu$ 
and finally vanishes at a critical value of $\mu=\mu_\mathrm{crit}$.  
In addition, the values of $m_0$, where the first order phase transition 
takes place, is shifted to larger values when $\mu$ is increased. 

The motivation of our investigation has been to provide a complementary
analysis of a ``QCD-like'' model to the findings of numerical simulations
\cite{Farchioni:2004us,Farchioni:2004ma,Farchioni:2004fs,Farchioni:2005tu} 
and from (Wilson) chiral perturbation theory
\cite{Sharpe:1998xm,Munster:2004am,Scorzato:2004da,Sharpe:2004ny,Sharpe:2004ps,Aoki:2004ta,Sharpe:2005rq}.
Both of these approaches have some shortcomings: in the numerical simulations
it is not clear whether the metastability effects are due to a bad 
behaviour of the algorithms employed. 
$\chi$PT on the other hand can, in principle, only predict the 
phase structure reliably 
in a region close to the continuum limit where the
$O(a^2)$ expansion is valid and the quark masses are small. 

Our findings from the large-$N$ analysis of the GN-model 
as described above show a striking similarity to the results from the 
four-dimensional simulations and from $\chi$PT. 
This includes the shift in the critical mass, the weakening of the 
first order phase transition as function of increasing $\mu$ 
and the vanishing of the first order phase transition altogether when 
$\mu>\mu_\mathrm{crit}$.
However, 
there is also a very noticeable and important difference to the picture we have
presently in four-dimensions: this concerns the coexistence of the first 
order phase transition with a second order transition to the Aoki phase. 
It is unclear to us whether this difference is an artefact of the 
GN-model in the large-$N$ analysis. Another, very interesting possibility is
of course that so far both, the numerical simulations and the analysis 
in $\chi$PT have missed this possibility. 
Clearly, if the Aoki phase transition is very close to the first order 
phase transition it will be very difficult for the numerical simulations
to resolve both. It could be speculated that the coexistence of both phases 
will show up in $\chi$PT only when higher orders are taken 
into account\footnote{Indeed, in some recent calculations in $\chi$PT for 
four-dimensional lattice-QCD, taking higher order effects 
into account such a coexistence of the Aoki phase and the first order 
phase transition --as we found here-- seems to take 
place \cite{Aokiprivate}.}. 
From our analysis here we would just like to give a warning that maybe the
phase structure of Wilson lattice QCD is even more complicated than 
our present picture suggests. 
If the picture of the phase diagram from our analysis in the GN-model 
would be correct also for lattice QCD, 
it would provide an argument for using 
a non-vanishing twisted mass parameter, since then one would avoid the
complication of the existence 
of two phase transitions.

We also want to mention that  
the $\sigma$-phase transition, 
which is present in the GN-model with Wilson fermion, 
changes to a normal first order phase transition,
when $\mu$ is switched on.

In this letter we did not address a number of additional interesting 
points that would deserve a further investigation:
in order to determine the phase transition, 
in particular in order to study the second order phase transition,
the second moment, for instance the susceptibility, may be better.
In principle one could also study the scaling behaviour of physical
quantities analytically toward the continuum limit.
In addition, an analysis within perturbation theory could provide
important insight into the renormalization properties of the GN-model 
with the twisted mass term.

Finally, it would be straightforward to change the discretization of the
fermions in the GN-model. For example, one could study a chiral invariant 
GN-model. In this way, it would be possible to learn how the change
of the fermion part of the action changes the phase structure. This is clearly
important for four-dimensional lattice QCD simulations.

\section*{Acknowledgments}
We thank Taku Izubuchi, Andrea Shindler and Giancarlo Rossi for 
very useful discussions and helpful communications.

\bibliographystyle{h-physrev4}
\bibliography{gn}

\end{document}